\documentstyle[preprint,aps,epsf,psfig]{revtex}
\begin{document} 
\title{Dynamics of Collapse  of  flexible  Polyelectrolytes and Polyampholytes}
\author{Namkyung Lee and  D. Thirumalai}
\address{IPST, 
University of Maryland, College Park, Maryland  20742} 
\date{\today} 
\maketitle 
 
\begin{abstract}
We provide a theory for the dynamics of collapse of strongly charged polyelectrolytes (PEs)
and flexible polyampholytes (PAs) using Langevin equation.
After the initial stage, in which counterions condense onto PE, the mechanism 
of approach to the globular state is similar for PE and PA.
In both instances, metastable pearl-necklace structures 
form in characteristic time scale that is proportional to  $N^{\frac 45}$
where $N$ is the number of monomers.  The late stage of collapse 
occurs by merger  of clusters  with the largest one growing at the expense of smaller ones (Lifshitz- Slyozov mechanism).  The time scale for this process $\tau_{COLL} \sim N$.  Simulations are used to support the proposed collapse 
mechanism for PA and PE.   
\end{abstract}
\tightenlines
Due to the interplay of several length scales charged polymers, 
referred to as polyelectrolytes (PEs), exhibit diverse structural 
characteristics depending on the environment 
(pH, temperature, ionic strength etc)\cite{barrat}.  The collapse  of 
polyelectrolytes  mediated by counterions is relevant    in  describing 
the  folding of  DNA and RNA.
Reversible condensation of DNA induced by multivalent
cations is required for its efficient 
packaging\cite{bloomfield1}.
Similarly, the role of divalent cations in enabling RNA to form its 
functionally competent state   has been firmly 
established\cite{Williamson}.  In both  cases,  a highly 
charged polyelectrolyte undergoes a collapse transition 
from an ensemble of  extended conformations.  

 Recently, simulations as well as theories 
have been advanced  to describe  
the collapse of flexible polyelectrolyte chains mediated by condensation 
of counterions\cite{Lee,micka,Solis,Olvera,Schiessel}.  
An understanding  of the collapse transition    in polyelectrolyte chains  could be 
valuable in gaining
insights into the condensation mechanism in biomolecules. 
Motivated by this we provide  a description of the  
dynamics of  collapse of flexible PEs and polyampholytes (PAs).  The theory
for the latter becomes possible because of our suggestion (see below) 
that the physical processes governing  the collapse of polyampholytes 
and polyelectrolytes are similar.  

In poor solvents (with reference to the uncharged polymer) the 
polyion  is stretched  at sufficiently high temperatures due to intramolecular 
electrostatic repulsion.
 The translational entropy of the counterions  at high 
temperatures is dominant, and hence their interaction with the 
polyion is negligible.  As the temperature is decreased the 
binding  energy of the  counterions  to the polyelectrolyte  
exceeds the free energy gain  due to  translational entropy
and they condense\cite{manning}.
 This results in a great decrease in the effective charge on the
 polyion and, because  the solvent is poor this leads to the compaction 
 of the chain.  

These considerations have been used by  Schiessel and Pincus\cite{Schiessel}
to propose a   diagram of states for highly charged polyelectrolytes.
Consider a PE consisting of $N$ monomers with a fraction $f(\approx 1)$ of monomers carrying a charge of 
$-e$. Strongly charged PE implies $f(l_B/A)^2 >1$ where 
 $l_B=e^2/4\pi \epsilon k_B T$ is the Bjerrum length, $\epsilon$ is  the 
dielectric constant of the solvent, and $A $ is the mean distance 
between charges.  
At low temperatures we expect counterions (with valence $z$)
  to condense onto the PE  because the 
Manning parameter\cite{manning} $(l_B/A) > 1/z$.
The electrostatic blob length $\xi_{el}$ is the scale at which the mutual 
repulsion between two charges is approximately $k_BT$, and is 
given by $\xi_{el}\approx l_B z^2/k^2$ where $k=-ln\phi$ 
with $\phi$ being the volume fraction of free counterions\cite{Schiessel}.
The size of the PE is stretched i.e. $L\approx k^2A^2N/l_Bz^2$ provided 
$\xi_{el}< \xi_{T} \approx a_o\Theta/(\Theta-T)$ where $\xi_{T}$ is the thermal blob  length, $\Theta$ is the collapse temperature for the uncharged polymer. When $\xi_{T} < \xi_{el}$ the PE undergoes a transition to a globular state.
   
 Here, we study the kinetics of this collapse 
process using Langevin equation.
In our earlier  study  we showed, using simulations, that the  approach to
 the globular state occurs in three stages following a temperature quench\cite{Lee}. 
 In the initial stage the counterions condense. 
The intermediate time regime is characterized by the formation of
metastable  necklace-pearl
structures\cite{Lee}.  In the final stage the pearls (domains) merge leading to the compact globular conformation.  
Here,  the theory for the intermediate time regime is
developed by adopting the procedure  suggested by Pitard and Orland\cite{Pitard}
to describe the dynamics of collapse (and swelling) of homopolymers.
We describe the dynamics of the late stages of collapse 
 using an analogy to Lifshitz-Slyozov\cite{LS} growth mechanism.

The equation of motion of the polyelectrolyte chain is assumed to be given  by the Langevin equation
\begin{equation}
\frac{\partial r}{\partial t} = - \frac{1}{\zeta} (\frac{\partial H }{\partial r}) + \eta(s,t)
\label{langevin}
 \end{equation}
where $\zeta = k_BT/D$ and $D$ is diffusion constant of monomer,
T is  the temperature.   The Hamiltonian of the  flexible 
polyelectrolyte chain with an effective Kuhn length $a_o$ is 
\begin{equation}
H(t) = \frac{k_BT}{2 a_o^2} \int ^N _0 ( \frac{\partial  r(s,t)}{\partial s})^2 ds +  V_c(t) 
\label{hamiltonian}
\end{equation}
and 
\begin{equation}
V_c(t) = \int^{N}_0ds\int^{N}_0ds' \frac{q(s)q(s')}{|\bf r\rm (s,t)-\bf r\rm (s',t)|} 
\label{Vc}
\end{equation}
The thermal noise $\eta(s,t)$ is assumed to be  Gaussian  with 
zero mean and the correlation is   
given by 
\begin{equation}
<\eta(s,t)\eta(s',t')> = 2D \delta(t-t').
\end{equation}
In Eq.(\ref{Vc}) $r(s,t)$ 
is the location of monomer $s$ at time $t$, $V_c$ is  Coulomb
interaction between monomers at $s$ and $s'$, and  $l_B$ is the Bjerrum length.
  In writing Eq.(\ref{Vc}) we have neglected hydrodynamic interactions\cite{Pitard,Pitard2}. 
Furthermore, excluded volume interactions are also omitted. 
The effects due to self-avoidance are not likely to be important in the early 
stages of collapse because in this time regime attractive interactions mediated by counterion condensation dominate. 

\it Collapse of Polyelectrolytes:\rm\,\,  
The polyelectrolyte chain is initially assumed to be  in the $\Theta$-solvent.  At $t=0$,
we imagine a quench to a temperature below $\Theta$ so that the chain is 
effectively in  a poor solvent. The equilibrium conformations of a 
weakly charged 
polyelectrolyte $(f(l_B/A)^2 \ll 1)$, in poor solvents has been described by Dobrynin et al\cite{dobrynin}. and has been  further illustrated by Micka et al\cite{micka}.  These studies showed that when the net charge 
on the polyelectrolyte chain exceeds  a certain critical value then the equilibrium conformation  resembles  a pearl-necklace.  This structure consists of clusters of charged droplets connected by strings under tension.  This is valid when counterion condensation, which is responsible for inducing  attraction between the monomers, is negligible.  

Consider a  charged polyelectrolyte chain $(f=1)$.   
Upon a  quench to $T < \Theta $, such that the thermal 
blob length $\xi_{T}$ is not too small, the polyelectrolyte chain undergoes a 
sequence  of structural changes enroute to  the collapsed conformation\cite{Lee}.
In particular, after counterion condensation PE evolves towards metastable pearl-necklace structure.  The dynamics of 
this process can be described using Eqs.(1-4) and the following physical picture. We assume that shortly following a quench to poor 
enough solvent conditions the counterions condense onto the polyelectrolyte chain.  
The time  scale  for condensation   of the multivalent counterions is 
diffusion limited\cite{Lee}, and is approximately given by 
$\tau_{COND}\approx \rho^{-2/3} \zeta/k_BT$ where $\rho$ is the density of the counterions.   
 Explicit simulations  reveal that the $\tau_{COND}$ 
 is much shorter than time scales in which   the 
  macromolecule relaxes.\cite{Lee}.   Upon condensation of a multivalent cation with 
 valence $z(\geq2)$, the effective charge around the monomer  
 becomes  $(z-1)e$.  If the locations of the divalent cation occur  randomly
and if the correlations between the  counterions are negligible   
 then in the early stages, PE chain with condensed counterions may be mapped onto an evolving  random 
 polyampholyte\cite{mapping}.  We  assume that the condensed counterions are 
 relatively immobile over the  time interval   
 $\tau_{COND} < t < \tau_{CLUST}$
 where $\tau_{CLUST}$ is 
 the time required for the formation of pearl-necklace structures.  
 With these assumptions  
the correlation between the renormalized 
charges on the polyanion can be written as 
\begin{equation}
 \overline{q_iq_j} =   q_o^2(N \delta_{ij}-1)/(N-1)
\label{correlation}
\end{equation}
 where $q_o$ is the net charge of the polyion after counterion condensation and  the average is done over the conformations of the polyelectrolyte with condensed counterions.

The above physical picture, which finds  support in explicit numerical
 simulations\cite{Lee}, can be used to calculate the dependence of the  early stage dynamics of the collapse process on $N$ using the method introduced by Pitard and Orland\cite{Pitard}.  
The basis of this method is to construct a reference Gaussian Hamiltonian  $H_v(t)$ with an effective Kuhn length $a(t)$
\begin{equation}
H_v(t)= \frac{k_BT}{2 a^2(t)} \int ^N _0 ( \frac{\partial r_v(s,t)}{\partial s})^2 ds.  
\label{hamiltonian2}
\end{equation}
  The value of $a(t)$  is determined from the condition that the difference in the size of the polyelectrolyte chain computed using  $H_v$ and the full theory vanish to first order in $\delta H = H_v-H $ for all $t$.
The equations of motion for the reference chain specified by $r_v(s,t)$ and 
for $\chi(s,t)= r(s,t)-r_v(s,t)$ are given (to first order in $\chi$) by 
\begin{eqnarray}
\frac{\partial r_v(s,t)}{\partial t} &=& - \frac{D}{ a^2(t)} (\frac{\partial^2 r_v(s,t) }{\partial s^2}) + \eta(s,t) \nonumber \\ 
\frac{\partial \chi(s,t)}{\partial t} &=& - \frac{D}{ a^2(t)} (\frac{\partial^2 \chi(s,t) }{\partial s^2}) + D[(\frac{1}{a_o^2}-\frac{1}{a^2(t)}) \frac{\partial ^2 r_v}{\partial s^2} + F(r_v(s,t))]
\label{eq-of-motion}
 \end{eqnarray}
where $ F(r(s,t))= -\partial( V_c/k_BT)/\partial r(s,t)$.
The   relaxation of  each mode is obtained in terms of Fourier representation
 $\tilde{r}_n(t)=1/N \int^{N}_{0} e^{i\omega_n s} r(s,t)ds$ $(\omega_{n}= \frac{2\pi n}{N})$  and is given  by
\begin{eqnarray}
\tilde{r}_n(t) &=& \int ^{t} dt_1 G_n(t-t_1)\eta_n(t_1) + G_n(t)\tilde{r}_n(0) \nonumber \\ 
\tilde{\chi}_n(t)&=& D\int^t dt_1 G_n(t-t_1) [F_n(r_v(t_1))- \omega_n^2(\frac{1}{a_o^2}-\frac{1}{a^2(t_1)})\tilde{r}_n(t_1)] 
\end{eqnarray}
where $ G_n(t) = exp(-D\int^{t} \frac{\omega_n^2}{a^2(t')}dt')$.

We assume that at $t=0$ the neutral chain  is  
in $\Theta$-solvent.
Therefore $<\tilde{r}_n(0) \tilde{r}_m^*(0)>= \frac{Na_o^2}{4 \pi^2 n^2}\delta_{mn}$.
 The thermal noise in Fourier space  satisfies
$<\tilde{\eta}_n(t)>=0$ and $<\tilde{\eta}_n(t)\tilde{\eta}^*_m(t')> = \frac{2D}{N} \delta_{nm}\delta(t-t')$.
We also assume that  $a(t)$ is a slowly varying 
 function  of time  for  $t\ll \tau_{ROUSE}$ where $\tau_{ROUSE}$ 
is defined implicitly by $\frac{4\pi^2D}{N^2} \int^{\tau_{ROUSE}}_0 \frac{d\tau}{a^2(t)} = 1$. 
 Thus,  we can let  $G_n(t) = exp(-D\omega_n^2t/a^2)$ for long wavelength modes.

The time dependent $R_g^2(t)$ can be expressed in terms of the amplitude 
and the relaxation rate of each mode
\begin{equation}
R_g^2(t) = \sum_n <|\tilde{r}_n^2(t)|> =\sum_n \frac{1}{N\omega_n^2}[(a_o^2-a^2(t)) G_n(t) +a^2(t)].
\label{rg2}
\end{equation}
The variational parameter $a(t)$ is obtained by equating the true value of the 
square of the radius of gyration $R_g^2$ to one calculated with $H_v$.  This implies that all the correction terms to $R_g^2$
 computed using Eq.(\ref{hamiltonian2}) should vanish.  To first order in $\chi(s,t)$
 the effective dynamic Kuhn length $a(t)$ satisfies the self consistent equation \cite{Pitard}
\begin{equation}
 \int^N_0 <r_v(s,t)\chi(s,t)>ds = 0
\end{equation}  
where the thermal average is done with respect to the noise term in the Langevin equation. 

The self consistent equation for $a(t)$ is 
\begin{equation}
(\frac{1}{a_o^2}-\frac{1}{a^2(t)})
= \frac{\sum_n\int^t dt' G_n(t-t') <\tilde{r}_n(t) \,\frac{\partial(\frac{- \Delta H}{k_BT})^*}{\partial \tilde{r}_n(t')}>}
{\sum_n \omega_n^2 \int^t dt' G_n(t-t')<\tilde{r}_n(t)\tilde{r}^*_n(t')>}
\label{self}
\end{equation}
where $\Delta H$ is the total interaction energy (Coulomb,excluded volume) 
of the PE. 
 
In the early  stage of the coil-to-globule transition 
 $t \ll \tau_{CLUST}$, we can assume $a^2(t) \approx a^2_o$.
The initial driving force for collapse is the counterion-mediated coulomb 
attraction. All other forces (such as hydrophobic interactions due to the 
poor solvent quality) play a less significant role. 
 Thus,  we can set $\Delta H = V_c$. With this observation the evaluation of $<r_n(t) F^*(r_n(t'))>$ gives 
\begin{equation} 
<\tilde{r}_n(t) F^*(r_n(t'))> \,\,
= \,\, < \tilde{r}_n(t) \tilde{r}^*_n(t')>\int \int  \frac{\sqrt{2}}{3\pi^2} \frac{<q(s)q(s')> c_n^2(s,s')}{(\Omega(s,s',t')/2)^{3/2}} dsds'
\label{self2}  
\end{equation}
with $c_n (s,s') = e^{ins}-e^{ins'}$ and  $ \Omega(s,s',t)  =  \frac{1}{3} \sum_l c_l^2(s,s') <\tilde{r}_l^2(t)> $.
The physical picture relating the state of the PE chain (till formation
of pearl-necklace structure) to polyampholyte  is used to perform an
additional  average 
 over the ``quenched'' random charge variables (see Eq.(\ref{correlation})).
The resulting equation when substituted   into  the right hand side of Eq.(\ref{self}) yields an  expression for the time dependent Kuhn length $a(t)$, namely,
\begin{equation}
a^2(t) = a_o^2 [ 1- c_1 \frac{(Dt)^{3/4}}{k_BT}(a_o^{-2/5}  l_B \frac{N}{N-1})]
= a_o^2(1- (\frac{t}{\tau_c})^{3/4})  
\label{a2}
\end{equation}
Since $\overline{qq'}=-q_o^2/(N-1) $ when $s\neq s'$, we obtain $\tau_c\approx (a_o/l_B)^{4/3} a_o^2/D  $ and it is nearly independent  of the  polymer size $N$.
Note that $\tau_c$ is an estimate for the time scale in which the clusters in 
 the metastable 
pearl-necklace structures form.  Because the formation of such clusters occur
locally (i.e. by  interaction  between  monomers that are not separately by a long contours) $\tau_c$ is  expected to be independent of $N$.
  Therefore, we identify    $\tau_{c} = \tau_{CLUST}$\cite{note}.
In the time range  $\tau_{COND}< t< \tau_{CLUST}$,  the  radius of gyration
is  obtained by substituting  Eq.(\ref{a2})  into Eq.(\ref{rg2}).
We find the decay of $R_g(t)$ can be  approximated as  
\begin{equation} 
 R_g^2(t) \approx R_g^2(0)(1 - (t/\tau_{PE})^{\alpha_{PE}})
\label{rg2t}
\end{equation}
with   $\alpha_{PE} =5/4$ and $ \tau_{PE}  =(\frac{\pi N a_o}{12\sqrt{ 2D}})^{\frac45}(\tau_{CLUST})^{\frac53} $. The characteristic time  $\tau_{PE}$ is therefore $\tau_{PE} \sim (Na_o/l_B)^{\frac45}(a_o/D)^{\frac75}$.

The formation of local clusters  has  also 
been suggested  as a mechanism for  collapse  of homopolymers in poor solvents.
\cite{kuz,gennes}
The time scale for cluster formation for homopolymer has been calculated 
by Pitard and Orland\cite{Pitard} who found $\tau_{HP}\sim (a_o^2/D)(a_o^3/|v_2|)^{4/3} N^{4/3}$.
The counterion-mediated attraction leading to pearl-necklace like structures 
occurs on a shorter time scale $\tau_{PE}\sim N^{4/5}$.
In very poor solvents ($l_B/a_o < |v_2|^2/a_o^6 $), where the  hydrophobic 
interactions  dominant  the driving force for  collapse of PE, we expect 
the dynamics to  resemble  that of homopolymer\cite{Lee}.  

\it Dynamics of PA collapse:\rm\,\,   The theory developed for PE is directly applicable to describe 
the dynamics of collapse of PA without having to justify the averaging \cite{mapping} over  the quenched random charges (cf. Eq.(\ref{correlation})). 
The Langevin dynamics for PA is given by Eqs.(1-4) where the quenched random 
charges explicitly satisfy Eq.(\ref{correlation}). 
It is known that when the total charge of the PA  $Q< \pm \sqrt{N}e$
the chain is collapsed at low temperature regardless of the quality of 
the solvent\cite{kantor}.  The mathematical description of the dynamics given by Eq.(\ref{rg2t}) describes the early stages of collapse of PA.  For PA, described by Eqs.(1-4), there is no counterion condensation.  The metastable pearl-necklace structure
forms after the random charges on the monomers are turned on provided 
$Q < \pm\sqrt{N}e$\cite{kantor}.
We predict that the time for 
forming the metastable pearl-necklace conformations is also given by Eq.(\ref{rg2t})
The  power law decrease in radius of gyration (Eq.(\ref{rg2t})) 
is limited only to time scale within which  the structure leads 
to  several clusters. 
 
\it  Late stages of collapse:\rm \,\, For both PA and PE, the pearl-necklace
 structures merge (see below)  at $t>\tau_{CLUST}$ to form compact collapsed
structures. This occurs by the largest cluster growing at the expense of
smaller ones (''packman'' effect) which is reminiscent of the
Lifshitz-Slyozov\cite{LS} description of the kinetics of precipitation in 
supersaturated solutions. The driving force for this growth is the
concentration gradient across the clusters or domains. If this analogy is
correct then we expect that the size of the largest cluster $S(t)$ to grow
as 
\begin{equation}
S(t)\sim t^\alpha 
\label{LS}
\end{equation}
with $\alpha \approx \frac 13$. The collapse is complete when $S(t)\sim
R_g(t\rightarrow \infty )\sim a_o N^{\frac 13}$ which implies that the
characteristic collapse time $\tau _{COLL}\sim N$.

\begin{figure}[t]
\leavevmode\centering\psfig{file=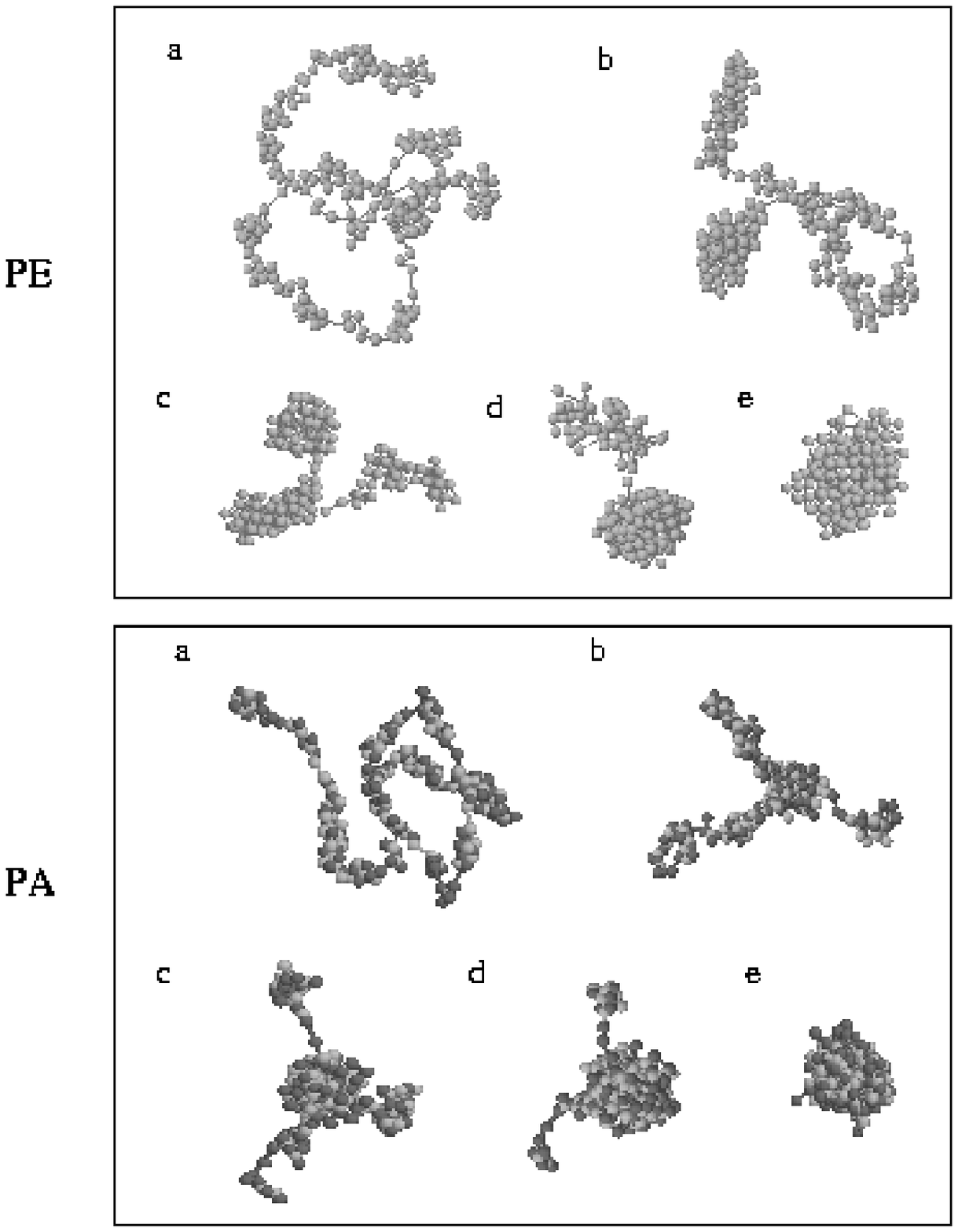,width= 12 cm}
\caption{
Snapshots of the conformations 
sampled in the dynamics of approach to the globular state starting from $
\Theta $-solvent conditions.
The final temperature is $l_B = 5 a_o$. For PA dark circles have 
$q = +e$ and grey circles have $q = -e$.  The simulations are performed by integrating the  Langevin equations.  The details are given Ref.[4]. 
 The top  panel  is for PE and
the lower one is for PA;   
$N=240$ for both PE and PA;  $v_2 = -2.97 a_o^3$ for PE  $v_2 = -6.18 a_o^3$ 
for PA  (the second virial coefficient $v_2$ gives an estimate of the 
quality of the solvent[4]). 
The counterions are explicitly included only in  the simulation of PE.
The solvent quality is such that the combined systems of the collapsed PE chain and the counterions form a Wigner Crystal[4].  The corresponding 
morphology of PA is clearly  more disordered (compare structure 'e' 
in the top and bottom panel).   
}
\label{fig2}
\end{figure}

In order to validate the proposed mechanism, i.e., the formation of
necklace-globule structures and the growth of the largest
domain by devouring the smaller ones, we performed Langevin simulations\cite{Lee} for
strongly charged flexible PEs and PAs in
sufficiently poor solvents so that the equilibrium structure is the compact
globule. In Fig.(\ref{fig2}) we display examples of the conformations that are
sampled in the dynamics of approach to the globular state starting from $
\Theta $-solvent conditions. Both panels (top is for PEs and
the lower one is for PAs) show that in the later stages of
collapse the largest clusters in the necklace-globule grow and the smaller
ones evaporate. This lends support to the proposed Lifshitz-Slyozov
mechanism. 
 
\begin{figure}[h]
\leavevmode\centering\psfig{file=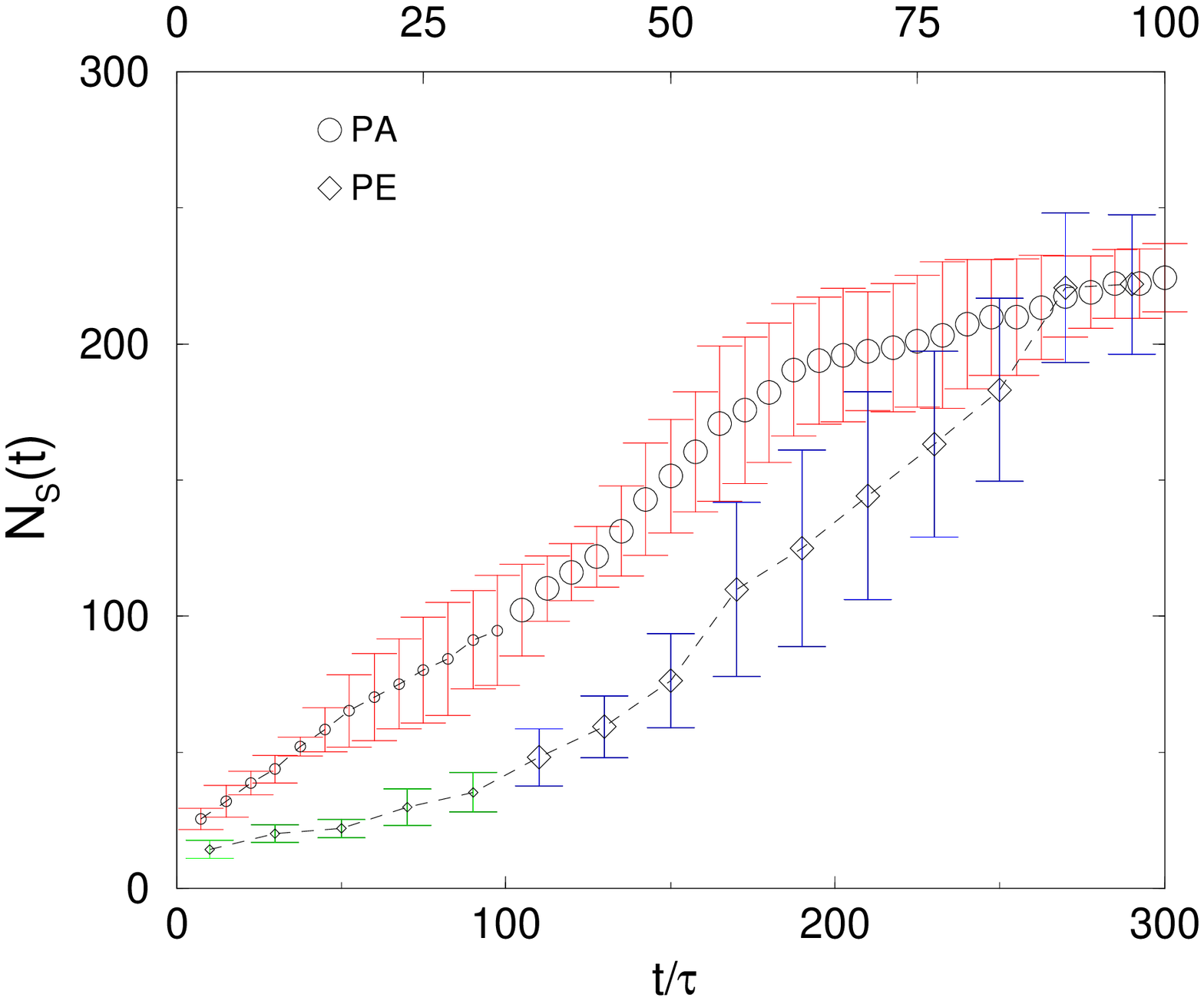,width= 11 cm}
\caption{
Time  evolution of the number of particles belonging  to the largest 
cluster $N_S(t)$ in a randomly charged single polyampholyte  and a 
polyelectrolyte ($N=240$).  The time scale for PA is given in the upper line. 
There are inherent numerical difficulties in  computing $N_s(t)$  for finite sized  systems.  For a given trajectory
there is not a sharp boundary for the largest cluster because of this the error-bars are large.  Nevertheless, we observe  linear increase of $N_S(t)$ 
for $t>100 \tau$ (PE) and $t > 30\tau$ for (PA) where $\tau = a_o^2/D$.  This  
implies  the Lifshitz-Slyozov growth mechanism
 i.e. $S(t) \sim N_S(t)^{1/3} \sim t^{1/3}$.  At long times $N_s(t)$ 
saturates indicating the completion of globule formation. 
}
\label{fig3}
\end{figure}

In order to estimate $\alpha$ (cf. Eq.(\ref{LS})) we have used  molecular dynamics simulations of polyelectrolyte and 
 randomly charged  polyampholyte  and 
 calculated    the number of particles that belong to the  largest cluster $N_S(t)$  as a function of time.
 In Fig.(\ref{fig3}), the simulation results show the linear increase of $N_S(t)$ for times greater than $t  \approx 30\tau$ for PA and $t\approx 100\tau$ 
 for PE.   Here $\tau= a_o^2/D$.   Therefore,    we  estimate  $\alpha \approx\frac 13$.
The change of slope for long times  is   due to the finite size 
effects and indicates the completion of the globule formation. Because the
quality of solvent (see legend in Fig.(\ref{fig3}) is not the same  for PA and PE the
onset time for linear behavior occurs at different times.

\it Conclusions:\rm\,\,
 We have presented a unified picture of collapse dynamics of PE and PA.
Although the morphology of collapsed structures for PA and PE are different,
our theory shows that the mechanism of approach to the globular state 
for both should be similar.  
In particular, both PA and PE reach the collapsed conformations via 
metastable pearl-necklace structures. For PE the driving force for forming
such structures is the counterion-mediated attractions. Charge fluctuations
in PA lead to pearl-necklace structures.
The life time of such structures is determined by a subtle competition
between attractive interactions mediated by counterions  and hydrophobic interactions determined by the solvent quality.  The interplay between 
these forces may be assessed  by estimating  the free energy of necklace-globule structures. 
The necklace-globule conformation consists of $n$ globules with nearly 
vanishing net charge that  are in local equilibrium.
The free energy of  $i^{th}$ globule is 
\begin{equation} 
 F_i \sim \frac{4\pi}{3} (\Delta f) R_i^3 +4\pi\sigma  R_i^2
\end{equation}
where $R_i$ is the radius of the $i^{th}$ globule and $\sigma$ is the surface tension.
 Note that  $\Delta f$ is the same  before and after the merger of clusters.
The free energy difference  between  a conformation consisting of two 
clusters and the conformation in which they are merged is $\Delta F \sim (8\pi\sigma - 4 (2)^{2/3}\pi\sigma) (N/n)^{2/3}$. 
Typical charge fluctuation in  each globule is $\sim q_o(N/n)^{1/2}$.
If the coulomb energy fluctuation  of each globule ($\delta E_{fluct} \sim 
q_o^2/a_o (N/n)^{2/3} $) is less  than the  free energy difference between
the conformation with two separated clusters and the one where they are merged,
then the   system  spontaneously grows to a  large cluster. 
If the solvent quality is not very poor, i.e.$ \sim (8\pi\sigma - 4 (2)^{2/3}\pi\sigma)<  q_o^2/a_o $, the life time of the metastable
 necklace-globule  can be long so that the collapse mechanism is controlled by the energy barrier between the  metastable state and the globule.
In this case, the attractive interaction between the clusters are induced by the mobile 
charge (counterion)  fluctuations\cite{ha2}
just as for  the  counterion mediated interaction between like charged rods.
The morphology of the final collapsed state is also determined by 
a competition  between electrostatic interactionis and 
hydrophobic forces.  When 
the solvent is very poor the collapse state is amorphorous whereas when 
collapse is determined by $\Delta F$  the globular state is an ordered 
Wigner Crystal\cite{Lee}.

\begin{references}
\bibitem{mapping} The mapping of PE with condensed counterions to PA rests on the assumption that
the condensed counterions are relatively immobile on time scales 
$(\tau_{COND}< t<\tau_{CLUST})$.  This is not expected to be strictly 
obeyed because because lateral motions (along the chain backbone) of 
counterions does occur.   
 We justify 
this by explicit comparisons between theory and Langevin simulations.  
 For PA the charge correlations are, by definition, given by Eq.(\ref{correlation}).
\bibitem{note} For homopolymers it follows from [13] that $\tau_{h} \sim 
a_o^6/v_2^4D$ where $v_2 \approx (\Theta -T/\Theta) a_o^3$.  This time scale 
may correspond to local cluster formation seen in computer simulations\cite{kuz}.
\end{references}
\end{document}